\documentclass[%
 reprint,
superscriptaddress,
 amsmath,amssymb,
 aps,
 prl,
floatfix,
]{revtex4-1}

\usepackage{url}
\usepackage{hyperref}
\usepackage{longtable}
\usepackage{bbm}
\usepackage{times}
\usepackage{nicefrac}
\usepackage{amsfonts}
\usepackage[ansinew]{inputenc}
\usepackage{epsfig}
\usepackage{graphicx}
\usepackage{color}
\usepackage{bm}
\usepackage{multirow}
\usepackage{mathrsfs}
\usepackage{revsymb}
\usepackage{amssymb}
\usepackage{amsmath}
\usepackage{mathtools}
\usepackage{slashed}
\usepackage{dcolumn}
\usepackage[babel]{csquotes}
\usepackage{lineno,hyperref}
\modulolinenumbers[5]
\usepackage{amsmath}
\usepackage{xspace}
\usepackage{graphicx}% Include figure files
\usepackage{dcolumn}% Align table columns on decimal point
\usepackage{bm}% bold math
\usepackage{subfigure}% bold math

\newcommand{\pentatrap}{\textsc{Pentatrap}\xspace}

\begin{document}

\title{Mass-difference measurements on heavy nuclides with at an eV/c$^2$ accuracy level with \pentatrap}

\author{A. Rischka}
\thanks{Present address: ARC Centre for Engineered Quantum Systems, School of Physics, The University of Sydney, NSW 2006, Australia.}
\email[]{alexander.rischka@mpi-hd.mpg.de}
\affiliation{Max-Planck-Institut f\"ur Kernphysik, 69117 Heidelberg, Germany}

\author{H. Cakir}
\author{M. Door}
\author{P. Filianin}
\author{Z. Harman}
\author{W.J. Huang}
\affiliation{Max-Planck-Institut f\"ur Kernphysik, 69117 Heidelberg, Germany}

\author{P. Indelicato}
\affiliation{Laboratoire Kastler Brossel, Sorbonne Universit\'e, CNRS, ENS-PSL Research University, Coll\`ege de France, Paris, France}

\author{C.H. Keitel}
\affiliation{Max-Planck-Institut f\"ur Kernphysik, 69117 Heidelberg, Germany}

\author{C.M. K\"onig}
\author{K. Kromer}
\author{M. M\"uller}
\affiliation{Ruprecht-Karls-Universit\"at Heidelberg, 69117 Heidelberg, Germany}

\author{Y. N. Novikov}
\affiliation{Department of Physics, St Petersburg State University, St Petersburg 198504, Russia}
\affiliation{Petersburg Nuclear Physics Institute, 188300 Gatchina, Russia}

\author{R. X. Sch\"ussler}
\author{Ch. Schweiger}

\author{S. Eliseev}
\author{K. Blaum}
\affiliation{Max-Planck-Institut f\"ur Kernphysik, 69117 Heidelberg, Germany}

\date{\today}

\begin{abstract}
First ever measurements  of the ratios of free cyclotron frequencies of heavy highly charged ions with Z $>$ 50 with relative uncertainties close to 10$^{-11}$ are presented. Such accurate measurements have become realistic due to the construction of the novel cryogenic multi-Penning-trap mass spectrometer \pentatrap. Based on the measured frequency ratios, the mass differences of five pairs of stable xenon isotopes, ranging from $^{126}$Xe to $^{134}$Xe, have been determined. Moreover, the first direct measurement of an electron binding energy in a heavy highly charged ion, namely of the 37$^{\text{th}}$ atomic electron in xenon, with an uncertainty of a few eV is demonstrated. The obtained value agrees with the calculated one  using two independent different implementations  of the multiconfiguration Dirac-Hartree-Fock method. \pentatrap opens the door to future measurements of electron binding energies in highly charged heavy ions for more stringent tests of bound-state quantum electrodynamics in strong electromagnetic fields and for an investigation of the manifestation of Light Dark Matter in isotopic chains of certain chemical elements.

\end{abstract}

\maketitle

Many areas of fundamental physics require the knowledge of mass differences or mass ratios of a variety of nuclides with very low uncertainty \cite{BLAUM20061,EdMyers2019}. Notable examples are, e.g., neutrino physics \cite{PTMS_for_Neutrinos}, a test of special relativity \cite{Rainville2005} and bound-state quantum electrodynamics (QED) \cite{kohler2016isotope}, ion clocks \cite{kozlov2018highly} and the search for Dark Matter via high-resolution isotope shift measurements \cite{PhysRevA.97.032510,Dark_Matter_1,Dark_Matter_2}.\\
In order to satisfy these requirements, the novel experiment \pentatrap \cite{repp2012pentatrap,roux2012trap}  has been set up at the Max Planck Institute for Nuclear Physics in Heidelberg. This is the first experiment that finally offers a realistic opportunity to pursue a mass-ratio measurement on heavy (A$>$100) highly charged ions with a fractional uncertainty below 10$^{-11}$.\\
\pentatrap is based on high-precision Penning-trap mass spectrometry (PTMS) - nowadays the only technique  which enables mass-ratio measurements with such low uncertainties \cite{EdMyers2019}. This arises from the fact that in the Penning trap one converts the determination of the mass $m$ of an ion with charge $q$ into a determination of the free cyclotron frequency $\nu_c = \frac{1}{2\pi}\frac{q}{m} B$ of the ion  stored in 
a combination of a strong uniform magnetic field $B$ and a weak harmonic electrostatic potential well. In practice, one measures the frequencies $\nu_-$, $\nu_z$, $\nu_+$ of three independent trap eigenmotions (magnetron, axial and cyclotron motions, respectively) which an ion undergoes in the Penning trap and applies the invariance theorem $\nu_c^2 = \nu_-^2 + \nu_z^2 + \nu_+^2$ \cite{brown1986geonium} to determine the free cyclotron frequency.\\ 

In this Letter we present the first mass measurements carried out with \pentatrap on stable isotopes of xenon. The mass differences of five pairs, namely $^{134}$Xe-$^{132}$Xe, $^{132}$Xe-$^{131}$Xe, $^{131}$Xe-$^{129}$Xe, $^{129}$Xe-$^{128}$Xe and $^{128}$Xe-$^{126}$Xe, have been determined with relative uncertainties of a few 10$^{-11}$ by measuring the ratios of free cyclotron frequencies of the corresponding xenon isotopes in a charge state of 17+. The stable xenon isotopes were chosen for a first operation of \pentatrap for three reasons. First, some of the stable xenon isotopes belong to a group of just a few nuclides for which mass ratios have been determined with sub 10$^{-10}$ relative uncertainty \cite{wang2017ame2016},

In fact the only other setup which can reach relative uncertainties of a few 10$^{-11}$ is the FSU-trap\cite{fsuxenon}, but only on low charged ion species.
e.g., the mass differences of $^{134}$Xe-$^{132}$Xe, $^{132}$Xe-$^{131}$Xe and $^{131}$Xe-$^{129}$Xe have been determined with Florida State University (FSU) trap \cite{fsuxenon} and thus can be considered a suitable candidate for reference measurements for \pentatrap. Second, xenon has many (virtually) stable isotopes and thus in this respect is similar to isotopic chains proposed in  \cite{PhysRevA.97.032510,Dark_Matter_1,Dark_Matter_2} for the search for Dark Matter.
Third, the addressed xenon isotopes span a mass range of eight atomic mass units forming a reference mass comb for on-line Penning-trap experiments in this mass region and thus serve as a backbone of precision mass measurements for the Atomic Mass Evaluation (AME) \cite{wang2017ame2016}.\\     
To demonstrate the capability of \pentatrap to perform measurements of mass ratios with a fractional uncertainty of close to 10$^{-11}$, we also determined the binding energy of the 37$^{\text{th}}$ electron in Xe by measuring the ratio of the free cyclotron frequencies of $^{131}$Xe$^{17+}$ and $^{131}$Xe$^{18+}$ ions. The obtained value is compared with the even more precisely known theoretical prediction published in this Letter. This "proof-of-principle" measurement opens the door to future measurements of electron binding energies in very highly charged heavy ions (e.g. in hydrogen-like xenon ions) with an uncertainty of an eV required to perform stringent tests of bound-state QED in strong electromagnetic fields \cite{haffner2000testing}. Moreover measuring the binding energy of many -- or even all -- shell electrons benchmarks atomic structure theory in a challenging and previously inaccessible way. \\

The design of the \pentatrap experiment rests upon several basic requirements that are prerequisite to high-precision mass measurements: (1) the use of highly charged ions to increase the sensitivity of the experiment since the cyclotron frequency $\nu_c$  scales with the charge state $q$, (2) the long storage of a single ion in an ultra-high vacuum in a very small volume by  cooling the ion's surrounding environment to the temperature of liquid helium and (3) the application of the fast phase-sensitive frequency-measurement techniques PnP (Pulse-and-Phase) \cite{PhysRevA.41.312} and PnA (Pulse-and-Amplify) \cite{PnA}.\\ 
 
\begin{figure*}[t]
\includegraphics[width=1\textwidth]{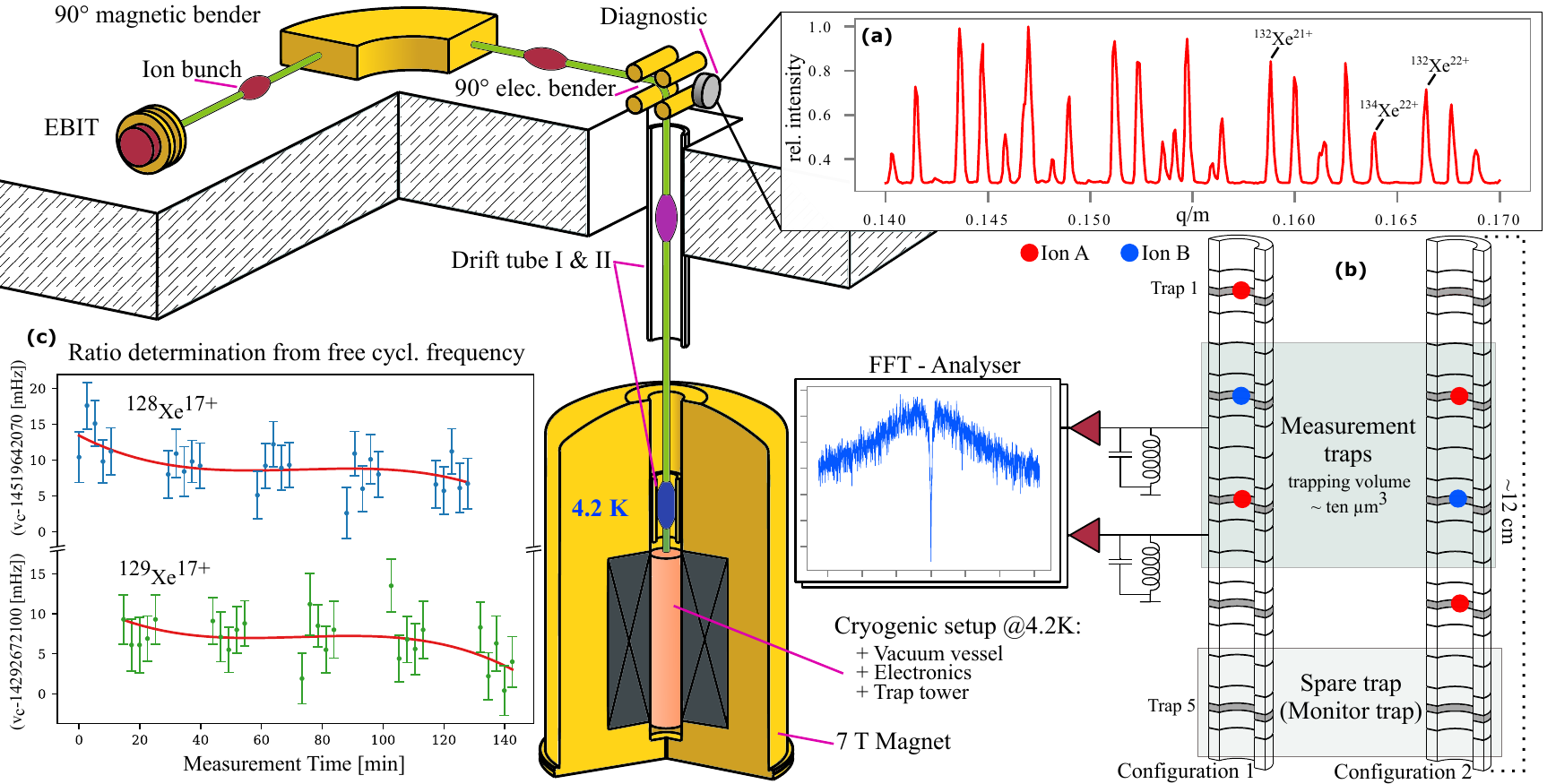}
\caption{(color online) Schematic overview of the \pentatrap setup and the measurement procedure. The setup contains an external ion source (EBIT) and a beamline in which the bunched ion beam is guided to the trap tower in the 7\,T magnet. (a) On the upper right a mass spectrum is shown which was recorded by scanning the magnetic bender. (b) The two trap towers represent the two different configurations in which the ions are stored during the measurement cycle. (c) On the lower left  a typical 2-hour measurement of the free cyclotron frequencies of $^{129}\textrm{Xe}^{17+}$ (green) and $^{128}\textrm{Xe}^{17+}$ (blue) ions is shown. The red line is a 3$^{rd}$-order simultaneous polynomial fit to the free cyclotron frequencies. For details see text.}
\label{fig:PENTATRAP_layout}
\end{figure*}

Highly charged ions of xenon are produced with a room-temperature Dresden-EBIT ion source \cite{zschornack2010status}. With a typical kinetic energy of 7 keV/q they are ejected as 1-$\mu$s long bunches from the EBIT and sent through a 90-degree sector dipole magnet, see Fig. \ref{fig:PENTATRAP_layout}. The sector magnet serves as a $q/m$ separator with a resolving power of about 400 allowing us to have a highly purified beam for many cases of our interest (see Fig. \ref{fig:PENTATRAP_layout}a). Prior to the injection of the chosen ions into the 7-\,T magnetic field of the mass spectrometer the kinetic energy of the ions is lowered in a room-temperature pulsed drift tube  from a few keV/q to about 200 eV/q. The final deceleration of the ions to nearly zero energy required for their capture in the traps takes place in  a second cryogenic pulsed drift tube, which is situated right above trap 1. The ions are captured in trap 1 by a reflecting potential well produced by the lowermost trap electrode and the restored high potential of the cryogenic pulsed drift tube. Adiabatic transportation of the ions to the other traps is realized by adiabatic changes of the potentials of the trap electrodes. The vacuum chamber that houses the traps and the associated cryogenic frequency-measurement electronics is cooled down to the temperature of liquid helium, 4.2 K. 
 
The mass spectrometer is located 
in a room with temperature stabilization to a few 10 mK/day. Together with a stabilization of the helium pressure and level in the magnet bore around the traps and active shielding of the trap region from stray magnetic fields this results in a very stable magnetic field with a relative time variation better than $\le 1 \cdot 10^{-10}/\text{h}$.\\         
 \indent Prior to the measurement of the ion trap frequencies in one of the measurement traps, the amplitudes of the ion motions are reduced to a few $\mu$m via the resistive cooling technique \cite{brown1986geonium}. The axial motion is cooled via its direct coupling to an LC-circuit which has a quality factor of 4000 (trap 2) or 9000 (trap 3) and a temperature below 10 K. The magnetron and cyclotron motions are cooled via their side-band coupling to the axial motion. The resistive cooling technique is also used to measure the frequencies of the trapped ion. In this case it is named "single-dip" method if the axial frequency is measured, and "double-dip" method in the case of the measurement of the magnetron or cyclotron frequencies \cite{feng1996tank}.  The technique of choice for the measurement of the cyclotron frequency is the PnP technique due to its higher precision compared to the double-dip method \cite{PhysRevA.41.312}.

It can be split up into two steps: (1) a measurement of its frequencies with moderate accuracy using the dip-techniques in order to identify the ion species and  (2) a main measurement of the ion frequencies with high accuracy.
\\

During the main measurement procedure the cyclotron and the axial frequencies are measured $simultaneously$. This measurement scheme, employed for the first time in our work, substantially reduces the uncertainty in the determination of the free cyclotron frequency. After a single ion is prepared in the measurement trap, its cyclotron motion is excited via a 5\,ms-long dipole radio frequency ($rf$) pulse to a radius of approximately 10 $\mu$m in order to fix the initial cyclotron-motion phase. During the phase-evolution time the cyclotron motion freely evolves and accumulates a certain phase. 
This waiting time is used to measure the axial frequency.
In order to measure the accumulated phase afterwards, the energy of the cyclotron motion is transferred to the axial motion via a 30 ms-long quadrupole $\pi$-pulse. This imprints the accumulated phase of the cyclotron motion to the phase of the axial motion, which is in turn measured with the FT-ICR technique. Such a measurement is performed for two phase-evolution times, usually $t_1$=0.1 s and $t_2$=40 s resulting in a measurement of two accumulated phases $\phi_1$ and $\phi_2$. This yields the cyclotron frequency $\nu_+$=$\frac{\phi_2-\phi_1+2\pi n}{2\pi (t_2-t_1)}$, where $n$ is the number of full cyclotron revolutions the ion performs in the time interval $t_2-t_1$. Since the magnetron frequency is small compared to the other ion frequencies, it is sufficient to measure it once a day during the preparatory step.\\
A unique feature of the \pentatrap mass spectrometer is its multi-trap configuration which consists of five identical cylindrical traps \cite{roux2012trap} (see Fig. \ref{fig:PENTATRAP_layout}b).
At the moment, two traps, trap 2, and trap 3, are used for measurements of the ion free cyclotron frequency, whereas the remaining three traps serve as ion storage containers. A measurement of the ratio of free cyclotron frequencies of two ions is performed with the following procedure. First, three ions of two species are loaded into trap 1, trap 2 and trap 3 (measurement configuration 1 in Fig.~\ref{fig:PENTATRAP_layout}b). The ions in trap 1 and trap 3 are identical. The free cyclotron frequencies of the ions in trap 2 and trap 3 are measured for approximately 15 min. Afterwards, the ion species in the measurement traps are swapped by adiabatically transporting them into the neighboring traps (measurement configuration 2 in Fig.~\ref{fig:PENTATRAP_layout}b) with subsequent measurements of their free cyclotron frequencies. This measurement procedure is then repeated.  
Such a multi-trap configuration significantly reduces the uncertainty in the mass-ratio determination. 
\\ 
The data analysis scheme of choice is called the "polynomial" method, which has already been successfully employed in several experiments \cite{FSU_POLY,eliseev2015direct, ISOLTRAP_POLY}. It is based on the reasonable assumption that the magnetic field drift and hence the free cyclotron frequency changes in time can be approximated by a polynomial of low order. 
Thus, the free cyclotron frequency drift of ion 1 and ion 2 is approximated by two polynomials that differ only by a coefficient of proportionality $R=\frac{\nu_c(ion1)}{\nu_c(ion2)}$. In the reported measurements on xenon isotopes we divided the measurement period into approximately 2-hour intervals and obtained $R$ for each interval from the simultaneous fit of two polynomials to the corresponding free-cyclotron-frequency values (see Fig. \ref{fig:PENTATRAP_layout}c). 
The interval duration is chosen so that on one hand the number of the frequency measurement points exceeds the number of free parameters of the polynomial, on the other hand the interval must be as short as possible to be able to approximate the free-cyclotron-frequency drift with a polynomial of a low order. Already a 2-hour measurement allows one to reach an uncertainty of 4 $\times$ 10$^{-11}$ in the determination of the free-cyclotron-frequency ratio of xenon ions in the 17+ charge state.  \\
\indent The mass differences of the five pairs of stable xenon isotopes were determined by measuring the free-cyclotron-frequency ratios of the corresponding xenon ions in a charge state of 17+. The results of the measurement are summarized in Table \ref{tab:massdiff}. Moreover, we also determined the binding energy of the 37$^{th}$ electron in xenon by measuring the ratio of the free cyclotron frequencies of $^{131}$Xe$^{18+}$ and $^{131}$Xe$^{17+}$ ions: $\nu_c(^{131}$Xe$^{18+})$/$\nu_c(^{131}$Xe$^{17+})$=1.058827929585(17)(30).  

Current experimental literature values for electron binding energies in highly charged ions of Xe-isotopes  are either not accurate enough or do not exist. Thus, we compare our experimental result with two theory values obtained from two independent groups, to be detailed in what follows.\\
The ground state of the $\null^{131}\textrm{Xe}^{18+}$ ion is a simple Kr-like configuration [Ar]$3d^{10} 4s^2 4p^6$.
The additional electron in Rb-like $\null^{131}\textrm{Xe}^{17+}$ is in the $4d_{3/2}$ state, i.e. the valence shell is filled according to Coulomb ordering instead of the usual
Madelung ordering.\\
Because of the high nuclear charge, relativistic corrections such as the Breit correction to the electron-electron is rather significant on the individual energy levels.
We apply the multiconfiguration Dirac-Hartree-Fock (MCDHF) method~\cite{Grant1970,Desclaux1971} to account for such relativistic correlation effects. Within this scheme, the many-electron atomic state function
is given as a linear superposition of configuration state functions (CSFs) sharing common total angular momentum ($J$), magnetic ($M$) and parity ($P$)
quantum numbers: $|\Gamma P J M\rangle = \sum_{k} c_k |\gamma_k P J M\rangle$.
The CSFs $|\gamma_k P J M\rangle$ are constructed as $jj$-coupled $N$-particle Slater determinants of one-electron wave functions, and $\gamma_k$ is a multi-index that
includes all the information needed to fully describe the CSF, i.e. orbital
occupation and coupling of one-electron angular momenta. $\Gamma$ collectively denotes all the
$\gamma_k$ included in the representation of the Kr- or Rb-like ground state. Using two different implementations of the MCDHF method~\cite{Indelicato2005,GRASP2018},
we systematically expand the active space of virtual orbitals to monitor the convergence of the calculations and to assess their uncertainties.\\
The binding energy difference for these two Xe ions is dominated by the Dirac-Hartree-Fock term.
Electron correlation effects contribute tens of eVs for both ions, however, since the $4d_{3/2}$ outermost valence electron polarizes the $3d$, $4s$ and $4p$ subshells, their
contribution is more complicated to account for in case of the Rb-like ion. Nevertheless, correlation terms largely cancel in the energy difference.\\
In highly charged ions, typically QED corrections such as the self-energy are of relevance.
The self-energy  corrections have been calculated in two different ways: the newly proposed
effective operator method from Refs.~\cite{Shabaev2015} has been compared to computing the Lamb shift of the $4d_{3/2}$ electron
(see~\cite{Yerokhin1999}) employing an effective radial potential which accounts for the screening by the core electrons.
However, the QED corrections amounting to approx. 21~meV are not observable at the current level of precision; the same thing can be said about the mass shift contributions.
One of the calculations employing the code in Ref.~\cite{Indelicato2005} uses full relaxation of all spectroscopic orbitals, and single and double excitations from all these orbitals to
the free single-electron states up to $5d$, with a result of 432.4(3.0)~eV. In the other computation, employing~\cite{GRASP2018}, we generate the set of CSFs with excitations from the $3s$--$4d$ states
up to $10h$, with the virtual orbitals optimized layer by layer, arriving to the value 435.1(1.0)~eV.\\
The measured value of 432.4(1.3)(3.4) agrees within one-sigma uncertainty with the theoretical values.
This measurement and theory comparison can be considered a "proof-of-principle" experiment for higher charge states where, e.g., stringent QED tests could be performed by measuring the binding energy of the remaining electrons in few-electron ions.
\\

\begin{figure}[t]
\includegraphics[width=0.5\textwidth]{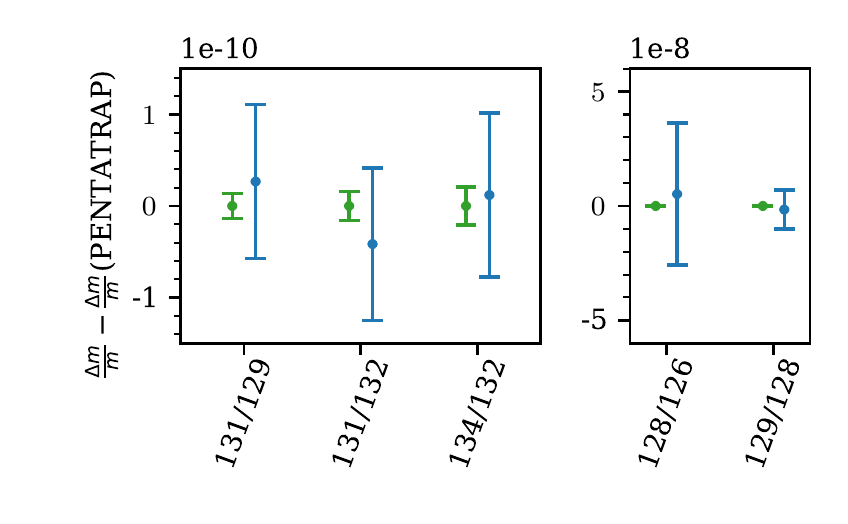}
\caption{(color online) Comparison of the relative mass differences $\Delta m/m$ of five stable xenon pairs determined in this work with the evaluated AME2016 values \cite{wang2017ame2016}. The green and blue error bars are the \pentatrap and AME2016 uncertainties, respectively. The zero point of the Y-axis corresponds to the mass difference of the corresponding xenon pair determined in this work.}
\label{fig:mass_diff}
\end{figure}

\begin{table*}[t]
    \caption{\label{tab:massdiff} Results of the measurements on the five stable xenon isotopes obtained in this work. The first and second columns list the addressed xenon ion pairs and their free cyclotron frequency ratios, respectively. The mass differences of the neutral states of these pairs determined on the basis of the measured free cyclotron frequency ratios are presented in the third column. The frequency ratios and the mass differences are given with their statistical and systematic uncertainties placed in the first and second round brackets, respectively. The mass differences evaluated in the AME2016 \cite{wang2017ame2016} are presented in the fourth column with their total uncertainties. The last column lists the so-called "improvement of accuracy" factor that demonstrates the increase of precision in the mass-difference determination achieved in this work compared to the AME2016 uncertainties. }
    \begin{ruledtabular}
    \begin{tabular}{llllc}
        ion pair                                  &    frequency ratio                  &   mass difference / u                   &   mass difference / u  & improvement      \\
                                                  &                                     &    (this work)                          &   (AME2016 \cite{wang2017ame2016})     & of accuracy    \\       
        \hline \\
        $^{134}$Xe$^{17+}$/$^{132}$Xe$^{17+}$   & 1.015~172~982~205(19)(8)              & 2.001~237~945~4(25)(11)                    &    2.001~237~947(12)       &  4   \\
        $^{132}$Xe$^{17+}$/$^{131}$Xe$^{17+}$   & 1.007~632~569~193(13)(6)              & 0.999~070~956~6(17)(8)                    &    0.999~070~951(11)       &  6   \\
        $^{131}$Xe$^{17+}$/$^{129}$Xe$^{17+}$   & 1.015~518~803~388(9)(8)               & 2.000~303~273~5(12)(10)                    &    2.000~303~277(11)       &  7    \\
        $^{129}$Xe$^{17+}$/$^{128}$Xe$^{17+}$   & 1.007~828~736~895(10)(6)               & 1.001~250~105~6(13)(8)                    &    1.001~249~9(11)         &  740 \\
        $^{128}$Xe$^{17+}$/$^{126}$Xe$^{17+}$   & 1.015~880~167~834(18)(8)              & 1.999~233~328~2(23)(10)                    &    1.999~234(4)           &  1700 \\  
    \end{tabular}
    \end{ruledtabular}
    \end{table*}
In order to determine the mass difference of neutral xenon isotope pairs from the measurement of the ratio of the free cyclotron frequencies of corresponding xenon isotope ions, we calculated the total binding energy of the 17 missing electrons to be 2982(5) eV. The obtained values of the mass differences for all five xenon pairs agree within 1-sigma with the values evaluated in the Atomic Mass Evaluation (AME2016) \cite{wang2017ame2016}. The uncertainty of the mass differences of three pairs, $^{134}$Xe-$^{132}$Xe, $^{132}$Xe-$^{131}$Xe, $^{131}$Xe-$^{129}$Xe \cite{fsuxenon}, were decreased by at least a factor of 4 (see Fig. \ref{fig:mass_diff} and Table \ref{tab:massdiff}). Whereas the mass differences of two pairs, $^{129}$Xe-$^{128}$Xe and $^{128}$Xe-$^{126}$Xe, were determined by at least two orders of magnitude more precisely.\\
\indent In the case of the mass differences the statistical and systematic uncertainties are similar in size, whereas the total uncertainty of the determination of the binding energy of the 37$^{\text{th}}$ electron in xenon is dominated by the systematic uncertainty. This is due to the fact that the  difference of the $m/q$-ratio in the case of $^{131}$Xe ions in the 17+ and 18+ charge states is by a factor of at least 3.5 larger than that of the different xenon isotopes in the same 17+ charge state. The total systematic uncertainty in the frequency ratio determination of $^{131}$Xe$^{18+}$/$^{131}$Xe$^{17+}$ is equal to 3$\times$10$^{-11}$ and is fully dominated by the impact of the linear gradient of the magnetic field along the trap axis on the values of the trap frequencies. The total systematic uncertainty in the frequency ratio determination of the xenon isotopes in the 17+ charge state does not exceed 10$^{-11}$ and is a result of  (1) the non-harmonicity of the trap electrostatic potential and of (2) the non-uniformity of the magnetic field.\\
\newline
\indent To conclude, we presented first ever measurements of the ratios of free cyclotron frequencies of heavy highly charged ions with relative uncertainties close to 1 $\cdot$ 10$^{-11}$. Based on the measured frequency ratios, we determined the mass differences of five pairs of stable xenon isotopes, $^{134}$Xe-$^{132}$Xe, $^{132}$Xe-$^{131}$Xe, $^{131}$Xe-$^{129}$Xe, $^{129}$Xe-$^{128}$Xe and $^{128}$Xe-$^{126}$Xe, with high precision, thus having demonstrated the ability of \pentatrap to perform high-precision measurements on an isotope chain of the same element. Such measurements on isotope chains such as Ca, Sr and Yb are of paramount importance to laser-spectroscopic experiments for Dark Matter searches \cite{PhysRevA.97.032510,Dark_Matter_1,Dark_Matter_2} and thus belong to planned measurements of the highest priority for \pentatrap.\\
Furthermore, the binding energy of the 37$^{\text{th}}$ atomic electron in xenon of 432.4(1.3)(3.4)\,eV was determined. This 'proof-of-principle' measurement finally renders realistic a way to stringently test QED in strong electromagnetic fields by comparing determinations of electron binding energies with $\leq 1$\,eV uncertainty in very highly charged heavy ions. With the current uncertainty of \pentatrap the measurement of the binding energy of the 11$^{\text{th}}$ atomic electron would be sufficient for a QED test.\\

We appreciate professional advice from Jos\'e R. Crespo L\'opez-Urrutia and Hendrik Bekker. This article comprises parts of the Ph.D. thesis work of H. C. and R. S.
This work is part of and supported by the German Research Foundation (DFG) Collaborative Research Centre ``SFB 1225 (ISOQUANT) and by the DFG Research UNIT FOR 2202.
P.I. acknowledges partial support from NIST. Laboratoire Kastler Brossel (LKB) is ``Unit\'e Mixte de Recherche de Sorbonne Universit\'e, de ENS-PSL Research University, du Coll\`ege
de France et du CNRS n$^\circ$ 8552''. P.I., Yu.N. and K.B. are members of the Allianz Program of the Helmholtz Association, contract n$^\circ$ EMMI HA-216 ``Extremes of Density and Temperature: Cosmic
Matter in the Laboratory''. P.I. wishes to thank Jean-Paul Desclaux for his help improving the MCDFGME code.
This project has received funding from the European Research Council (ERC) under the European Union's Horizon 2020 research and innovation programme under grant agreement No. 832848 - FunI. Furthermore, we acknowledge funding and support by  the International Max Planck Research School for Precision Tests of Fundamental Symmetries (IMPRS-PTFS) and by
 the Max Planck PTB RIKEN Center for Time, Constants and Fundamental Symmetries.

\bibliography{xenon_article}

\end{document}